\documentclass{llncs}
\usepackage[utf8]{inputenc}
\usepackage{amsmath}
\usepackage{graphicx}
\usepackage{amsfonts}
\usepackage{amssymb}
\usepackage{url}

\usepackage{booktabs}

\begin{document}

{\let\thefootnote\relax\footnotetext{(*)\emph{version as accepted at the BioASQ Workshop at CLEF 2020}~\\Copyright \textcopyright\ 2020 for this paper by its authors. Use permitted under Creative Commons License Attribution 4.0 International (CC BY 4.0). CLEF 2020, 22-25 September 2020, Thessaloniki, Greece.}}

\title{CoLe and LYS at BioASQ MESINESP8 Task: similarity based descriptor assignment in Spanish$^{(*)}$}
\titlerunning{CoLe and LYS at BioASQ MESINESP8}  % abbreviated title (for running head)
\author{Francisco J. Ribadas-Pena\inst{1}, Shuyuan Cao\inst{1}, Elmurod	Kuriyozov\inst{2}}
%\author{XXX1 XXXX XXXXX \inst{1}, YYY1 YYYY YYYYY\inst{2},  XXX2 XXXX XXXXX\inst{1}}
%elmurod
\authorrunning{Ribadas, Cao, Elmurod} % abbreviated author list (for running head)
%\tocauthor{Francisco J. Ribadas-Pena, Victor M. Darriba-Bilbao,  xxx, yyy}
%
%\institute{Departamento de Inform\'atica, Universidade de Vigo\\
%E.S. Enxe\~ner\'{i}a Inform\'atica, Edificio Polit\'ecnico, \\ Campus As Lagoas, s/n, 32004 Ourense (Spain) \\
%\email{\{ribadas,darriba\}@uvigo.es}
%}

\institute{
Grupo COLE, 
Departamento de Inform\'atica, Universidade de Vigo\\
E.S. Enxe\~nar\'{i}a Inform\'atica, Campus As Lagoas, Ourense 32004, Spain\\
\url{ribadas@uvigo.es}, \url{shuyuan.cao@uvigo.es}
\and
Grupo LYS, Departamento de Computaci\'on y Tecnolog\'{\i}as de la Informaci\'on, Universidade de A Coru\~na\\
Facultade de Informatica, Campus de Elvi\~na, A Coru\~na 15071, Spain
\url{e.kuriyozov@udc.es}           
}

\maketitle              % typeset the title of the contribution

\begin{abstract}
In this paper, we describe our participation in the {\sc mesinesp} Task of the BioASQ biomedical semantic indexing challenge. 
The participating system follows an approach based solely on conventional information 
retrieval tools. We have evaluated various alternatives for extracting index terms from IBECS/LILACS documents in order to be stored in an Apache Lucene index. Those indexed representations are queried using the contents of the article to be annotated and a ranked list of candidate labels is created from the retrieved documents. We also have evaluated a sort of limited Label Powerset approach which creates meta-labels joining pairs of DeCS labels with high co-occurrence scores, and an alternative method based on label profile matching. Results obtained in official runs seem to confirm the suitability of this approach for languages like Spanish.
\end{abstract}

\section{Introduction}
This article describes the joint participation of the CoLe group~\footnote{\emph{Compiler and Languages} group, \url{https://www.grupocole.org/}}  from the
University of Vigo and the LYS group~\footnote{\emph{Language in the Information Society} group, \url{https://www.grupolys.org/}} from the University of A Coru\~na
in the Spanish biomedical semantic indexing task of the 2020 BioASQ challenge~\cite{bioasq}. 
Participants in this task are asked  to automatically classify abstracts written in Spanish from two medical databases, IBECS and LILACS, 
labeling those documents with descriptors taken from the DeCS (\emph{Descriptores en Ciencias de la Salud}) structured vocabulary. 

%Both groups (CoLe~\footnote{\emph{Compiler and Languages} group, \url{http://www.grupocole.org/}} 
%from University of Vigo and 
%UTAI~\footnote{\emph{Uncertainty Treatment in Artificial Intelligence} group, \url{http://decsai.ugr.es/utai/}} %from University of Granada) 
%have participated in the previous BioASQ editions. Our previous participations 
%assessed the use of two different machine learning based techniques: a top-down arrangement of local classifiers %and 
%a Bayesian network induced by the thesaurus structure. Both 
%approaches modelled the task of assigning descriptors from the MeSH 
%hierarchy to MEDLINE documents as a hierarchical multi-label classification problem.

In our participation we have followed a similarity based strategy, where the
final list of DeCS descriptors assigned to a given article is created from the set of most similar IBECS/LILACS articles stored in a textual index created from the training dataset. This neighbor based strategy was explored in previous participations in BioASQ challenge~\cite{self1}, where we tested the suitability of this 
similarity based approach and evaluated several strategies to improve the final ranked list of descriptors.

In the case of text categorization for Spanish written documents in {\sc mesinesp} Task we have employed this similarity based method using several index term extraction approaches in order to evaluate the effects of document representation in the overal quality of the predicted labels. We have also tried improving the categorization performance using a sort of limited Label Powerset multi-label categorization approach, where meta-labels created by joining pairs of labels with high co-occurrence scores replace the original document labels. Additionally a similarity based method using synthetic documents to represent ''label profiles'' was evaluated and integrated into our official {\sc mesinesp8} runs.

The rest of the paper is organized as follows. Section 2 describes the main
ideas behind the proposed similarity based approach for {\sc mesinesp8} annotations and also
describes the text processing being applied. 
Section 3 briefly details the use of synthetic meta-labels in our Label Powerset approach and how we use the ''label profiles'' to annotate {\sc mesinesp} articles. 
Finally, section 4 discusses our official runs in the BioASQ challenge and details the most 
relevant conclusions of our  participation.

\section{Similarity based descriptor selection}\label{sec:knn}

Approaches based on $k$ nearest neighbors ($k$-NN) have been widely used in the context of large scale multi-label categorization, being employed for 
MEDLINE documents~\cite{MESHUP} and for labeling purposes in many other domains. The choosing of $k$-NN based methods is mainly due to its scalability, minimum parameter tuning requirements and,  despite its simplicity, its ability to deliver acceptable results in cases where large amounts of examples are available.
The approach we have followed in our BioASQ challenge participation~\footnote{Source code available at \url{https://github.com/fribadas/mesinesp8}.} is essentially a large multi-label $k$-NN classifier backed by 
an Apache Lucene~\footnote{\url{https://lucene.apache.org/}} index.
In the case of {\sc mesinesp} annotation with DeCS descriptors, despite being a complex problem, with more than 33,703 labels in DeCS 2019
arranged in a hierarchical structure, the availability of a fairly large training set ($>$ 318K abstracts) labeled by human experts, a priori supposes 
a favorable scenario for this $k$-NN labeling.

In this way, our annotation scheme starts by indexing the contents of the {\sc mesinesp} training articles. For each new article to annotate 
that index is queried using its contents as query terms. The list of similar articles returned by the indexing engine and their 
corresponding similarity measures are exploited to determine the following results:
\begin{itemize}
\item predicted number of descriptors to be assigned
\item ranked list of predicted DeCS descriptors
\end{itemize}

The first aspect conforms a regression problem, which aims to predict the number of descriptors to be included in 
the final list, depending on the number of descriptors assigned to the most similar articles identified by the indexing 
engine and on their respective similarity scores.
The other task is a multi-label classification problem, which aims to predict a descriptors list based 
on the descriptors manually assigned to the most similar {\sc mesinesp} articles.
In both cases, regression and multi-label classification, similarity scores calculated by the indexing engine are exploited. 
These scores are computed during the query processing phase. Query terms employed to retrieve the similar articles are extracted from the 
original article contents and linked using a global OR operator to conform the final query sent to the indexing engine.

In our case, the scores provided by the indexing engine are similarity measures resulting from the engine internal computations and the 
weighting scheme being employed, which do not have an uniform and predictable upper bound.
In order for these similarity scores to behave like a real distance metric, we have applied the following normalization procedure:
\begin{enumerate}
 \item Articles to be annotated are preprocessed in the same way than the training articles indexed by the Lucene engine.
 \item In classification time, all of the relevant index terms from the article being annotated  are joined by an 
       OR operator to create the search query.
 \item In the ranking of similar articles returned by the indexing engine the top result will be the same article used to query the index, this result 
       is discarded but its score value ($score_{\mbox{\tiny \sc max}}$) is recorded for future normalization.
 \item For each element on the remaining articles set, the number of descriptors is recorded and it is also recorded 
       the list of assigned descriptors, 
       %assigning to each of them a weight 
       %equals to $\frac{1}{\left (1-\frac{score}{score_{\mbox{\tiny \sc max}}}  \right ) ^2}$, which will be employed
       %in the weighted voting scheme of the $k$-NN classification.
       linking to each of them an estimated distance to the article being annotated,
       equals to $\left (1-\frac{score}{score_{\mbox{\tiny \sc max}}}  \right )$, which will be employed
       in the weighted voting scheme during  $k$-NN classification.
\end{enumerate}

With this information the number of descriptors to be assigned to the article being annotated is predicted
using a weighted average scheme, where the weight of each similar article is the inverse of the square of the estimated distance 
to the article being annotated, that is, $\frac{1}{\left (1-\frac{score}{score_{\mbox{\tiny \sc max}}}  \right ) ^2}$.

To create the ranked list of descriptors a distance weighted voting scheme is employed, 
associating the same weight values (the inverse of squared estimated distances) to the respective similar articles. 
Since this is actually a multi-label categorization task, there are
as many voting tasks as candidate descriptors were extracted from the articles retrieved
by the indexing engine. 
For each candidate label, positive votes come from similar articles annotated with it and negative votes come from articles not including it.

\begin{table}[th]
%\begin{center}
\caption{\label{table:test_weighting_k} Performance comparison of term extraction approaches.}
\scriptsize

\setlength{\tabcolsep}{8pt}
\begin{center}
\begin{tabular}{c}

\begin{tabular}{rrccccccc}	
\toprule
\                 & \bf k & \bf MiF & \bf MiP & \bf MiR & \bf EBF & \bf EBP & \bf EBR & \bf Acc \\   
\midrule
stems & 10 & 0.3241     & 0.3342 &    0.3145 &     0.3098 &     0.3391 & 0.3130 & 0.1968 \\
      & 20 & 0.3473     & 0.3586 &    0.3367 &     0.3319     & 0.3617 & 0.3360 & 0.2131 \\
      & 30 & 0.3517     & 0.3634 &    0.3407 &     0.3356     & 0.3656 & 0.3391 & 0.2155 \\
      & 40 & \bf 0.3569     & \bf 0.3679 &    0.3465 &     \bf 0.3404 & \bf 0.3691 & 0.3444 & \bf 0.2189 \\

\midrule
lemmas & 10 & 0.2635     & 0.2704     &     0.2569 & 0.2485 & 0.2737 & 0.2520 & 0.1532 \\
       & 20 & 0.2891     & 0.2974     &     0.2812 & 0.2719 & 0.2994 & 0.2729 & 0.1703 \\
       & 30 & 0.2988    &  0.3081 &         0.2901 & 0.2806 & 0.3090 & 0.2805 &  0.1765 \\
       & 40 & 0.2964     & 0.3057     &     0.2876 & 0.2777 & 0.3054 & 0.2774 & 0.1745 \\

\midrule
NPs & 10 & 0.2839 & 0.2899 & 0.2781 & 0.2685 & 0.2915 & 0.2733 & 0.1666 \\
    & 20 & 0.3079 & 0.3154 & 0.3008 & 0.2911 & 0.3150 & 0.2976 & 0.1823 \\
    & 30 & 0.3121 & 0.3201 & 0.3044 & 0.2948 & 0.3206 & 0.2992 & 0.1852 \\
    & 40 & 0.3156 & 0.3237 & 0.3080 & 0.2982 & 0.3237 & 0.3024 & 0.1880 \\

\midrule
DEPs & 10 & 0.1794 & 0.1917 & 0.1687 & 0.1635 & 0.1892 & 0.1576 & 0.0980 \\
     & 20 & 0.1982 & 0.2119 & 0.1861 & 0.1801 & 0.2078 & 0.1724 & 0.1092 \\
     & 30 & 0.2069 & 0.2206 & 0.1948 & 0.1876 & 0.2162 & 0.1794 & 0.1140 \\
     & 40 & 0.2087 & 0.2224 & 0.1966 & 0.1894 & 0.2174 & 0.1812 & 0.1151 \\

\midrule
all & 10 & 0.3247 & 0.3352 & 0.3148 & 0.3110 & 0.3389 & 0.3179 & 0.1967 \\
    & 20 & 0.3499 & 0.3612 & 0.3393 & 0.3351 & 0.3645 & 0.3416 & 0.2151 \\
    & 30 & 0.3536 & 0.3640 & 0.3437 & 0.3386 & 0.3662 & 0.3455 & 0.2172 \\
    & 40 & 0.3528 & 0.3634 & 0.3428 & 0.3373 & 0.3648 & 0.3439 & 0.2167 \\

\midrule
UIMA & - & 0.2305 & 0.1677 & \bf 0.3687 & 0.2475 & 0.2063 & \bf 0.3821 & 0.1487\\

\bottomrule
\end{tabular}	
	
\end{tabular}	

\end{center}

\end{table}

\subsection{Evaluation of article representations}\label{sec:representations}

In our preliminary experiments we have tested several approaches to extract the set of index terms to represent {\sc mesinesp} articles
in the indexing process.
%We have also evaluated the effects in annotation performance of the different weighting schemes available in the Apache Lucene indexing engine.

Regarding article representation we have evaluated four index term extraction approaches. In these experiments and also in the 
official {\sc mesinesp8} runs we have worked only with the Pre-processed Training set provided by BioASQ organizers. We have discards the 
dataset of PubMed abstracts translated into Spanish provided by {\sc mesinesp} organizers due to format issues regarding part of the available translated abstracts.
The final training dataset comprised 318,658 records with at least one DeCS code. Index terms which occurred in 5 or less articles were discarded and terms which were present in more than 50 \% of training documents were also removed.

Our  aim with these experiments was to determine whether linguistic motivated index term extraction could help to improve annotation performance
in the $k$-NN based method we have described. We employed the following methods:
\begin{description}
\item [Stemming based representation.] This was the simplest approach which employs stop-word removal, using a standard stop-word list for Spanish,
and the  default Spanish stemmer from the Snowball project\footnote{http://snowball.tartarus.org}. 

\item [Morphosyntactic based representation.] In order to deal with the effects of morphosyntactic variation in Spanish
we have employed a lemmatizer to identify lexical roots instead of using word stems and we also replaced stop-word removal 
with a content-word selection procedure based on part-of-speech (PoS) tags.

\hspace*{12pt} We have delegated the linguistic processing tasks to the tools provided by the spaCy Natural Language Processing (NLP) toolkit~\footnote{Available at \url{https://spacy.io/}}. This toolkit offers a set of state-of-the-art components written in the Python programming language, together with a collection of pretrained models, 
ready to be used in typical natural language processing tasks like dependency parsing, named entity recognition, PoS tagging and morphological analysis.

\hspace*{12pt} In our case we have employed the PoS tagging and lemmatization information provided by spaCy to tokenize and assign PoS tags to the
{\sc mesinesp} abstract contents. We employed the standard Spanish models available on spaCy without using any specific data for biomedical related contents. 

\hspace*{12pt} In order to filter the content-words from the processed {\sc mesinesp} abstracts, we have applied a simple selection criteria based on 
the employment of the PoS that are considered to carry the sentence meaning. Only tokens tagged as a noun, verb, adjective, adverb or as 
unknown words are taken into account to constitute the final article representation. 

\hspace*{12pt} After PoS filtering, the lemmas (canonical forms of words) corresponding to surviving tokens are employed 
to normalize the considered word forms in a slightly more consistent way than simple stemming.

\item [Nominal phrases based representation.] In order to evaluate the contribution of more powerful NLP techniques,
we have employed a surface parsing approach to identify syntactic motivated nominal phrases from which meaningful multi-word index 
terms could be extracted. 

\hspace*{12pt} Noun Phrase (NP) chunks identified by spaCy are selected and the lemmas of the constituent tokens are joined together to create a multi-word index term.
In the current version of the system no other syntactical units of interests like prepositional phrases or verbal phrases are considered, since nominal phrases 
use to carry most of the text semantic content.

\item [Dependencies based representation.] We have also employed as index terms triples of dependence-head-modifier extracted by the dependency parser provided by spaCy.
A dependency parser analyzes the grammatical structure of a sentence, establishing relationships between head words and words which modify those heads. In our case spaCy provides
a dependency parsing model for Spanish that identify syntactic dependency labels following the Universal Dependencies(UD)~\cite{UD} scheme.

\hspace*{12pt} Dependence relationships encode information that provides an approximation to high level semantic relationships, giving information regarding the agent of an action (with a \emph{nsubj} relationship between the main verb and the root of the nominal phrase acting as subject) or the object of that action (by means of a \verb!obj! relationship), among others. In our system, the complex index terms were extracted from the following UD relationships~\footnote{Detailed list of UD relationships available at \url{https://universaldependencies.org/u/dep/}}: \emph{acl}, \emph{advcl}, \emph{advmod}, \emph{amod}, \emph{ccomp}, \emph{compound}, \emph{conj}, \emph{csuj}, \emph{dep}, \emph{flat}, \emph{iobj}, \emph{nmod} , \emph{nsubj}, \emph{obj}, \emph{xcomp}, \emph{dobj} and \emph{pobj}.
 
\item [UIMA Concept Mapper representation.] In addition to those representations, we also have employed the Concept Mapper~\footnote{http://uima.apache.org/d/uima-addons-current/ConceptMapper/RELEASE\_NOTES.html} module from the UIMA (\emph{Unstructured Information Management Architecture}) framework. This component employs a dictionary with all of the DeCS labels and their corresponding synonyms and searches for exact matches of those DeCS labels into the abstract text. In our case we have added to the document representation as index term each one of those matches in order to maintain its absolute occurrence frequency. 

\end{description}

In order to illustrate the index term extraction procedure, figure~\ref{fig:example} shows an example of a {\sc mesinesp} record with the set of representations extracted from the textual contents of its abstract.

Table~\ref{table:test_weighting_k} summarizes the results obtained in our preliminary tests regarding document representation, using as measures MicroPrecision, MicroRecall, MicroF,
Example based Precison, Recall and F measure, and Accuracy. The test dataset was created after processing 750 manually indexed records from the \emph{Core-descriptors development set} provided by BioASQ organizers.
We have evaluated the performance of the described index term generation methods (\verb!stems!, \verb!lemmas!, \verb!nps!, \verb!deps! and using all of them together) for increasing values of $k$, the number of similar articles
to be used (1) in the estimation of the number descriptors to be assigned and (2) in the voting procedure that will construct the final list of 
descriptors to attach to a given article.
%
%We have also evaluated the effect of two index term weighting methods available in
%version 4.10 of Apache Lucene: a classical \emph{tf-idf} weighting scheme~\cite{tfidf} 
%and a more complex one inspired by the Okapi BM25 
%family of scoring formulae~\cite{bm25}.
%These weighting schemes are employed by the Lucene engine to compute the similarity scores used  
%to create the ranking of documents relevant to a given query. In our case, the query terms are all of the index terms
%extracted from the article to be annotated using one of the methods described before.

As can be seen in table~\ref{table:test_weighting_k}, the best results were obtained with fairly high values for $k$ ($\ge$ 30). Regarding the index term representations,
the runs which employed index term extracted by means of stemming (both stemming alone and stems mixed with the other index terms) provided the best performance.
The representations using complex index terms extracted from noun phrase chunks and dependencies triples offered poor performance, maybe because of
very infrequent index terms that can have the undesired effect of boosting internal scores in schemes
where inverse document frequencies are taken into account.

\begin{figure}[ht]
{\tiny \begin{tabular}{ll}
\toprule
\bf \sc id & biblio-1000005 \\
\bf \sc db & LILACS \\
\bf \sc journal  &  Oncol. (Guayaquil) \\
\bf \sc title    &  Manejo de Tumores de Mediastino, Serie de Casos \\
\bf \sc abstract &  Introducción: A pesar del difícil acceso anatómico para los tumores de mediastino, la resección  \\ 
                 & quirúrgica sigue siendo el mejor enfoque   diagnóstico y terapéutico. En la presente serie de casos  \\
                 & presentamos la experiencia de un centro oncológico en el abordaje de tumores del mediastino \\
                 & y sus resultados. \\
             &  Métodos: En el departamento de Jefatura de Cirugía Oncológica del Instituto Oncológico\\
             &  nacional de Solca-Guayaquil, durante los meses de Enero del 2013 a Enero 2017 se realizó \\
             &  un estudio descriptivo, retrospectivo. Se analizaron todos los casos de pacientes derivado \\
             & ...\\
\bf \sc decs codes & 9562,8650,21034,24375,21044,20174,14341,238,9062,21030,23039\\
\midrule
\sc stems &  manej,tumor,mediastin,seri,cas,introduccion,dificil,acces,anatom,tumor,mediastin,\\
                    & reseccion,quirurg,enfoqu,diagnost,terapeut,objet,present,seri,cas,present,experient,\\
                    & centr,oncolog,abordaj,tumor,mediastin,result,metod,departament,jefatur,cirug,oncolog,\\
                    & ... \\
%                    & \bf lemmas \\
\sc lemmas                    &  tumores,mediastino,serie,casos,difícil,anatómico,quirúrgica,mejor,diagnóstico,terapéutico,\\
                    &  presente,presentamos,oncológico,jefatura,cirugía,oncológica,instituto,oncológico,nacional, \\
                    &  solca,guayaquil,enero,enero,realizó,descriptivo,retrospectivo,analizaron,derivados,inicial,\\
                    &  ...  \\  
%                    & \bf NPs \\
\sc NPs                    &  manejo de tumores,mediastino,introducción,acceso,los tumores,mediastino,la resección,\\
                    &  el mejor enfoque,el objetivo,la presente serie,la experiencia,un centro,el abordaje,tumores,\\
                    &  mediastino,sus resultados,métodos,el departamento,jefatura de cirugía oncológica,\\
                    &  instituto oncológico nacional de solca,guayaquil,los meses,enero 2017,un estudio,los casos,\\
                    &  ... \\
%                    & \bf DEPs \\  
\sc DEPs                    &  nmod(tumores,mediastino),conj(diagnóstico,terapéutico),flat(jefatura,cirugía),\\
                    &  flat(cirugía,oncológica),flat(instituto,oncológico),flat(instituto,nacional), \\
                    & flat(instituto,solca),flat(instituto,guayaquil),advcl(analizar,previo),obj(previo,marcador),\\
                    & amod(marcador,tumoral),flat(tomografía,tórax),conj(analizar,realizar),nsubj(estudiar,variable),\\
                    & ...\\
%                    & \bf UIMA  \\ 
\sc UIMA                    &  14294,14294,9562,9562,9562,9562,9562,9562,9562,9562,9562,9562,9562,9562,9562,9562,9562,\\
                    & 9562,9562,9562,9562,8650,8650,8650,8650,8650,8650,8650,8650,8650,8650,8650,8650,8741,8741,\\
                    &   16771,28219,28219,1731,1731,1731,1731,24373,28632,28632,28599,28599,23274,23274,23484,23484,\\
                    &   ...\\                    
\bottomrule
\end{tabular}
}

\caption{\label{fig:example} Example of the index term extraction methods.}
\end{figure}

\section{Exploiting DeCS labels}\label{sec:candidateprocessing}

In this section we describe two approaches that try to improve labeling performance taking advantage of the information inherent to
DeCS labels. Even DeCS is a fairly large concept hierarchy we have tested the suitability of extending the label space using an approach inspired by
the Label Powerset(LP) method employed in multi-label categorization. In this case we create a set of ''meta-labels'' that replace pairs of DeCS 
labels which tend to appear together in training dataset.
The other aspect regarding DeCS labels that we have explored is to exploit the idea of ''label profiles''. These profiles represent the concepts behind each DeCS label
by means of a synthetic document that aggregates the contents of all of the abstracts annotated with a given label.

\begin{table}[th]
	%\begin{center}
	\caption{\label{table:metalabel} Most frequent codes in original dataset and when ''meta-labels'' are applied.}
	
\begin{center}
	\setlength{\tabcolsep}{5pt}

\begin{tabular}{c}
{\scriptsize Original MESINESP Dataset (number of codes: 23199)} \\

{\tiny 
\begin{tabular}{crcl}
\toprule
\bf rank & \bf freq & \bf code & \bf label \\
\midrule
1	& 225535	& 21034 & ''seres humanos'' (\emph{''humans''})\\
2	& 120433	& 21030 & ''femenino''      (\emph{''female''}) \\
3	& 105182	& 21044 & ''masculino''     (\emph{''male''}) \\
4	& 56161	& 331   & ''adulto''        (\emph{''adult''}) \\
5	& 41571	& 9062  & ''persona de mediana edad'' (\emph{''middle aged''}) \\
6	& 33115	& 29315 & ''adolescente'' (\emph{''adolescent''}) \\
7	& 27275	& 2694  & ''niño''    (\emph{''child''}) \\
8	& 15795	& 20174 & ''anciano'' (\emph{''aged''}) \\
9	& 14514	& 28612 & ''factores de riesgo'' (\emph{''risk factors''}) \\
10	& 14325	& 2715  & ''niño preescolar''  (\emph{''preschool child''}) \\
11	& 12408	& 7399  & ''lactante'' (\emph{''infant''}) \\
12	& 11121	& 22226 & ''recién nacido'' (\emph{''newborn''}) \\
13	& 11038	& 28611 & ''estudios retrospectivos'' (\emph{''retrospective studies''}) \\
14	& 10654	& 28596 & ''estudios transversales'' (\emph{''cross-sectional studies''}) \\
15	& 10567	& 841   & ''animales'' (\emph{''animals''}) \\
\bottomrule
\end{tabular}
}
~\\
~\\

{\scriptsize Meta-labels with NPMI threshold at 0.25 (number of codes: 50771) }\\

{\tiny
\begin{tabular}{crcl}
\toprule
\bf rank & \bf freq & \bf code & \bf label\\
\midrule
1	& 116576	& 21030.21034 & ''femenino''$\wedge$''seres humanos'' (\emph{''female''$\wedge$''humans''}) \\
2	& 102586	& 21034.21044 & ''seres humanos''$\wedge$''masculino'' (\emph{''humans''$\wedge$''male''}) \\
3	& 89717	& 21034       & ''seres humanos''  (\emph{''humans''}) \\
4	& 84759	& 21030.21044 & ''femenino''$\wedge$''masculino'' (\emph{''female''$\wedge$''male''}) \\
5	& 44104	& 331.21030   & ''adulto''$\wedge$''femenino'' (\emph{''adult''$\wedge$''female''}) \\
6	& 37698	& 331.21044   & ''adulto''$\wedge$''masculino'' (\emph{''adult''$\wedge$''male''}) \\
7	& 34076	& 9062.21030  & ''persona de mediana edad''$\wedge$''femenino'' (\emph{''middle aged''$\wedge$''female''}) \\
8	& 31766	& 9062.21044  & ''persona de mediana edad''$\wedge$''masculino'' (\emph{''middle aged''$\wedge$''male''}) \\
9	& 27602	& 331.9062    & ''adulto''$\wedge$''persona de mediana edad'' (\emph{''adult''$\wedge$''middle aged''}) \\
10	& 25381	& 21030.29315 & ''femenino''$\wedge$''adolescente'' (\emph{''female''$\wedge$''adolescent''}) \\
11	& 22792	& 21044.29315 & ''masculino''$\wedge$''adolescente'' (\emph{''male''$\wedge$''adolescent''}) \\
12	& 16576	& 331.29315   & ''adulto''$\wedge$''adolescente'' (\emph{''adult''$\wedge$''adolescent''}) \\
13	& 11438	& 28612       & ''factores de riesgo'' (\emph{''risk factors''}) \\ 
14	& 11344	& 9062.29315  & ''persona de mediana edad''$\wedge$''adolesente'' (\emph{''middle aged''$\wedge$''adolescent''}) \\
15	& 11044	& 2694.29315  & ''niño''$\wedge$''adolescente'' (\emph{''child''$\wedge$''adolescent''}) \\
\bottomrule
\end{tabular}
}
\end{tabular}
\end{center}
\end{table}

\subsection{Limited Label Powerset}
Label Powerset(LP)~\cite{lp1}~\cite{lp2} is a problem transformation approach to multi-label classification that seeks to convert a multi-label classification problem into a multi-class classification problem. The LP transformation creates one multi-class classifier trained on all unique label combinations found in the training data. This approach is unfeasible in the case of DeCS labeling due the large amount of different labels in the hierarchy: 33,703 labels in DeCS 2019, of which 23,197 are actually present in the training dataset.

Our approach limits classical LP multi-label categorization to the cases where only combinations of highly correlated pairs of labels are taken into account. To select the pairs of labels to be joined we have
computed the Normalized Pointwise Mutual Information (NPMI) between each pair of DeCS labels, $l_i$ and $l_j$, across the training dataset employing the following formula:
$$
NPMI(l_i, l_j) = \frac{PMI(l_i, l_j)}{-log(P(l_i, l_j))}
$$
Where $PMI$ is the Pointwise Mutual Information computed by:
$$
PMI (l_j, l_j) = log(\frac{P(l_i,l_j)}{P(i_i) \cdot P(l_j)})
$$
And where $P(l_i, l_i)$ is computed as $\frac{|\mbox{\tiny docs. labeled with $l_i$ and $l_j$}|}{|\mbox{\tiny docs. in training collection}|}$ and $P(l)$ is computed
as $\frac{|\mbox{\tiny docs. labeled with $l$}|}{|\mbox{\tiny docs. in training collection}|}$.

The measure $NPMI(l_i, l_j)$ normalizes the values of PMI  in $[-1, 1]$, resulting in -1 for a pair of labels never occurring together, 0 for independence, and +1 for complete co-occurrence of labels $l_i$ and $l_j$. 

In our experiments we have evaluated three thresholds (0.25, 0.50 and 0.75) to create new ''meta-labels'' joining pairs of labels whose NPMI scores are over them. Table~\ref{table:metalabel}
compares the most frequent codes in the original dataset and when ''meta-labels'' with NPMI scores above 0.25 replace the original codes.

Once these ''meta-labels'' are identified
we create new training documents replacing in the set of DeCS labels associated to each record the two original labels with the corresponding new ''meta-label''. 
To annotate the test articles we apply the  $k$-NN procedure described in previous sections over the new training documents where ''meta-labels'' were placed.

\subsection{Label profiles}

Another approach that we have tested in order to capture the semantics of the DeCS labels is the use of ''label profiles'' that try to represent the contents associated to each DeCS label and match incoming test documents to those profiles.

To create those DeCS label profiles we have followed a very simple approach that is easily integrated into our Lucene backed $k$-NN multi-label categorization scheme. 
\begin{enumerate}
	\item For each DeCS label we collect the index terms extracted from the abstracts of documents annotated with that label.
	\item With those lists of index terms we create a synthetic Lucene document concatenating the terms to create a big document that holds the representation of the ''label profile'' for the corresponding label.
	\item All those synthetic documents representing ''label profiles'' for every DeCS label are indexed into a Lucene index.
\end{enumerate}

To annotate an incoming article abstract text is processed as described in precedent section to extract its index terms. With those index terms the Lucene index of ''label profiles'' is queried and the top most similar synthetic documents are recorded to annotate that article with their corresponding labels.
The idea behind this approach is to improve the main $k$-NN annotation procedure, which follows a content-based method, with a complementary method focused on the labels and its semantic aspects.

\begin{table}[ht]
%\begin{center}
\caption{\label{table:bioasq} Official results for BioASQ  {\sc mesinesp8} Task.}
\scriptsize
\begin{center}
	\setlength{\tabcolsep}{3pt}
\begin{tabular}{@{} cccccccccccc @{}} 
\toprule		
system    & rank       & MiF        & EBP   & EBR   & EBF   & MaP   & MaR   & MaF   & MiP   & MiR   & Acc. \\  
\midrule
%---
best     & 1/25        &	0.4254	&0.4382	&0.4343	&0.4240	&0.3989	&0.3380	&0.3194	&0.4374	&0.4140	&0.2786 \\
iria-mix & 8/25        &	0.3892	&0.5375	&0.3207	&0.3906	&0.5539	&0.2263	&0.2318	&0.5353	&0.3057	&0.2530 \\
iria-1   & 10/25       &    0.3630	&0.5055	&0.2980	&0.3643	&0.5257	&0.1908	&0.1957	&0.5024	&0.2842	&0.2326 \\
iria-3   & 11/25       &	0.3460	&0.5432	&0.2674	&0.3467	&0.5789	&0.1617	&0.1690	&0.5375	&0.2551	&0.2193 \\
iria-2   & 12/24       &	0.3423	&0.4699	&0.2837	&0.3408	&0.4996	&0.1715	&0.1719	&0.4590	&0.2729	&0.2145 \\
iria-4   & 14/25       &	0.2743	&0.3070	&0.2635	&0.2760	&0.2655	&0.2925	&0.2619	&0.3068	&0.2481	&0.1662 \\
BioASQ\_Baseline& 15/25 &	0.2695	&0.2681	&0.3239	&0.2754	&0.3733	&0.3220	&0.2816	&0.2337	&0.3182	&0.1659 \\
\bottomrule 
\end{tabular}
\end{center}

%\end{center}  
\end{table}

\section{Official MESINEPS8	runs and discussion}

Although we have tested several alternatives to try to improve the results obtained by the Lucene based $k$-NN method, only the most
simple ones have been submitted to the official batches of BioASQ challenge. 

In table~\ref{table:bioasq} the official performance measures obtained by our runs in the {\sc mesinesp8} Task are shown.
The official runs sent during our participation were created using the following configurations.

\begin{description}
\item[iria1.] This run created the representation of {\sc mesinesp} articles using all of the index term extraction methods described in section~\ref{sec:representations}.
              During indexing and querying, terms appearing in 5 or less abstracts and terms used in more than 50\% of
              total documents were discarded. The number of neighbors used by the $k$-NN classifier is 30 and the predicted number of descriptors to be returned was increased a 10\% in order to ensure slightly better values in recall related measures.
              
\item[iria2.] For this run the same setup as \verb!iria1! was employed, but instead of using the original train dataset this runs employed the limited Label Powerset approach and indexed a new training dataset annotated with ''metalabels'' created by joining pairs of DeCS labels with a NPMI scores above 0.25.

\item[iria3.] This run was simply the intersection of the labels predicted by \verb!iria1! and \verb!iria2!.
              
\item[iria4.] This run created a set of ''label profiles'' over the train dataset employed in \verb!iria2!, that is, documents annotated with ''metalabels'' created by joining
              pairs of labels with NPMI scores over 0.25. In this case the number of labels to predict was fixed to 10 and the number of neighbors used by the $k$-NN classifier was 15.

\item[iria-mix.] This run was based on run \verb!iria1!, adding the predictions of \verb!iria4! and the exact matches provided by UIMA Concept Mapper. 

Labels predicted by \verb!iria4! but discarded by \verb!iria1! were added to the final list of candidate labels. The same procedure was applied to add the exact matches identified by means of Concept Mapper but not predicted by neither of \verb!iria1! and \verb!iria4!.
\end{description}

The results of our participation in the {\sc mesinesp8} task of the BioASQ biomedical semantic indexing challenge
were not far from the results of the most competitive teams, showing that similarity based methods can still be considered for large scale indexing tasks. As positive aspects of our participation, we have confirmed that $k$-NN methods  backed by conventional textual indexers like Lucene are a viable alternative for this kind of large scale problems, with minimal computational requirements and fairly good results in the case of Spanish biomedical abstracts.
We have also conducted a comprehensive evaluation of the performance of several alternatives to index term extraction, ranging from
simple ones, based on stemming rules, to more complex ones were natural language processing was required. 

The future lines of work are related with the improvement of natural language processing. In this participation we have employed general domain NLP models. Biomedical documents have many specific characteristics that suggest that custom NLP models trained with text from this domain will help to improve the performance of our classifier. 
Likewise, the use of "meta-tags" in this work opens a future line of research on the exploitation of the semantics inherent to the co-occurrence of tags.

% \section{Conclusions and future work}

%\vspace*{-4pt}
% \section*{Acknowledgements}
% Research reported in this paper has been partially funded by 
% ''Ministerio de Econom\'{\i}a y Competitividad'' and {\sc feder} (under projects  FFI2014-51978-C2-1 and TIN2013-42741-P)
% and by the Autonomous Government of Galicia (under projects R2014/029 and R2014/034).


\section*{Acknowledgements}
F.J. Ribadas-Pena and S. Cao have been supported by the Spanish Ministry
of Economy, Industry and Competitiveness (MINECO) through the ANSWER-ASAP project (TIN2017-85160-C2-2-R), 
and by the Galician Regional Government (Xunta de Galicia) under project ED431D 2017/12.

\noindent E. Kuriyozov received funding from the ANSWER-ASAP project (TIN2017-85160-C2-1-R) from MINECO,
and from Xunta de Galicia (ED431B 2017/01, ED431G 2019/01). He is also funded for his PhD by El-Yurt-Umidi
Foundation under the Cabinet of Ministers of the Republic of Uzbekistan.

%\vspace*{-6pt}

\end{document}